# Hardware Implementation of TDES Crypto System with On Chip Verification in FPGA

Prasun Ghosal, Malabika Biswas, Manish Biswas

**Abstract**—Security issues are playing dominant role in today's high speed communication systems. A fast and compact FPGA based implementation of the Data Encryption Standard (DES) and Triple DES algorithm is presented in this paper that is widely used in cryptography for securing the Internet traffic in modern day communication systems. The design of the digital cryptographic circuit was implemented in a Vertex 5 series (XCVLX5110T) target device with the use of VHDL as the hardware description language. In order to confirm the expected behavior of these algorithms, the proposed design was extensively simulated, synthesized for different FPGA devices both in Spartan and Virtex series from Xilinx viz. Spartan 3, Spartan 3AN, Virtex 5, Virtex E device families. The novelty and contribution of this work is in three folds: (i) Extensive simulation and synthesis of the proposed design targeted for various FPGA devices, (ii) Complete hardware implementation of encryption and decryption algorithms onto Virtex 5 series device (XCVLX5110T) based FPGA boards and, (iii) Generation of ICON and VIO core for the design and on chip verification and analyzing using Chipscope Pro. The experimental as well as implementation results compared to the implementations reported so far are quite encouraging.

——————————— ◆ ———————————

## 1 INTRODUCTION

BEYOND any doubt, the need for secure storage or transfer of information is an inextricable part of human history. Nowadays, the rapid evolution of communication systems offers, to a very large percentage of population, access to a huge amount of information and a variety of means to use in order to exchange personal data. Therefore, every single transmitted bit of information needs to be processed into an unrecognizable form in order to be secured. This enciphering of the data is necessary to take place in real time and for this procedure cryptography is the main mechanism to secure digital information. Due to the heavy increase in the volume of information data, a variety of encryption algorithms have been developed [1-7]. Among the different cryptographic algorithms, the most popular example in the field of symmetric ciphers is the Data Encryption Standard (DES) algorithm, which was developed by IBM in the mid-seventies.

The DES algorithm is popular and in wide use today because it is still reasonably secure and fast [2][5-7][9-11]. There is no feasible way to break DES, however because DES is only a 64-bit (eight characters) block cipher, an exhaustive search of 255 steps on average, can retrieve the key used in the encryption. A much more secure version of DES called Triple-DES (TDES), which is essentially equivalent to using DES three times on plaintext with three different keys. Naturally, it is three times slower than the original form of DES but it is way more secure.

This paper examines the full procedure of implementing a DES and Triple DES algorithm using a high-level hardware description language, VHDL, combined with the usage of FPGA technology. The complete design was synthesized for various FPGA devices of Spartan and Virtex series, viz Spartan 3, Spartan 3AN, Virtex E, Virtex 5 etc. The design is implemented and verified on a Virtex 5 FPGA development board from Xilinx using device XCVLX5110T. Next, ICON and VIO type of core was developed and the total implementation of DES and TDES was verified on chip using Chipscope Pro. The rest of the paper is organized as follows: Section II describes previous works and implementation of Cryptography algorithm i.e. DES and Triple DES, also give a brief introduction to the Data Encryption Standard and Triple Data Encryption Standard algorithm respectively. Section III will give experimental framework and results. Section IV will give the conclusions. An initial version of this work has been reported in [15].

## 2 BACKGROUND

### 2.1 Previous work

A lot of research & development are going on over DES & Triple DES. DES and Triple-DES are already implemented in Spartan –II devices [6]. The design and implementation of DES and TDES processors was reported in [7][12]. DES is also developed using the Handel-C. Others are reported in [3][13][14].

————————————————

- *Prasun Ghosal is with the Department of Information Technology, Bengal Engineering and Science University, Shibpur, Howrah, WB 711103, INDIA.*
- *Malabika Biswas is with the Department of Information Technology, Bengal Engineering and Science University, Shibpur, Howrah, WB 711103, INDIA.*
- *Manish Biswas is with the Department of Information Technology, Bengal Engineering and Science University, Shibpur, Howrah, WB 711103, INDIA.*

## 2.2 Data Encryption Standard Encryption

Complete function of DES algorithm can be described briefly as follows [1] [11]. DES is a block cipher. It operates on blocks of 64-bits in size. A 64-bit input block of plaintext will be encrypted into a 64-bit output block of cipher text. It is a symmetric algorithm, which means the same algorithm and key are used for encryption and decryption. The security of DES rests in the 56-bit key. The DES algorithm functions as follows [1-7] [11] [20].

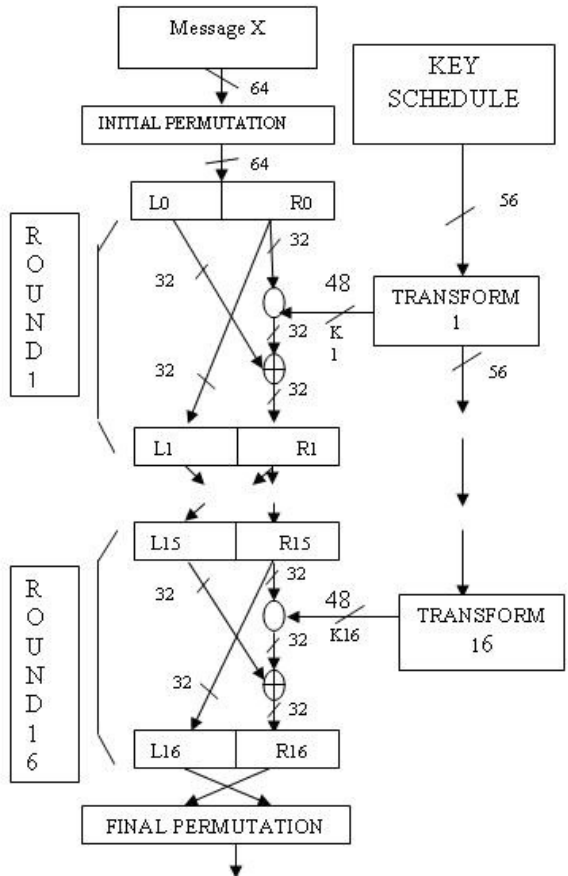

Fig 1: DES Encryption

The plaintext block is taken in and put through an initial permutation. The key is also taken in at the same time. The key is presented in a 64-bit block with every 8th bit being a parity check. The 56-bit key is then extracted ready for use. The 64-bit plaintext block is split into two 32- bit halves, named the right half and left half. The two halves of the plaintext are then combined with data from the key in an operation called Function F. There are 16 rounds of Function f, after which the two halves are recombined into one 64-bit block, which is then put through a final permutation to complete the operation of the algorithm and a 64-bit cipher text block is outputted. The detailed procedure is omitted due to paucity of space and is represented with a flowchart in Figure 1.

## 2.3 Data Encryption Standard Decryption

The decryption method is similar and omitted due to the paucity of the space.

## 2.4 Triple Data Encryption Standard

A concise representation of Triple Data Encryption Algorithm is described.
TDES is a block cipher operating on 64-bit data blocks. There are several forms, each of which uses the DES cipher three times. TDES can however work with one, two or three 56-bit keys. This means that the plaintext is, in effect, encrypted three times [2][5-7][9-11]. A number of modes of TDES have been proposed:

• DES-EEE3: Three DES encryptions with three different keys.
• DES-EDE3: Three DES operations in the sequence encrypt-decrypt-encrypt with three different keys.
• DES-EEE2 and DES-EDE2: Same as the previous formats except that the first and third operations use the same key.

Let $E_K(I)$ and $D_K(I)$ represent the DES encryption and decryption of I using DES key K respectively. Each TDEA encryption/decryption operation is a compound operation of DES encryption and decryption operations. The following operations are used:
1. TDEA encryption operation: the transformation of a 64-bit block I into a 64-bit block O that is defined as follows: $O = E_{K3}(D_{K2}(E_{K1}(I)))$.
2. TDEA decryption operation: the transformation of a 64-bit block I into a 64-bit block O that is defined as follows: $O = D_{K1}(E_{K2}(D_{K3}(I)))$.

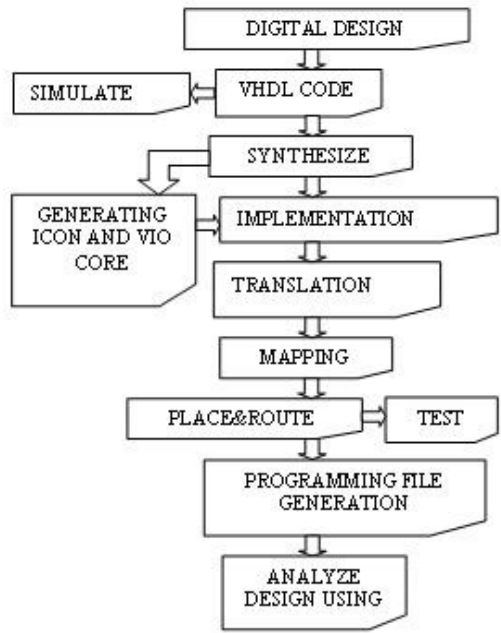

Fig 2: Different steps in course of implementation

## 3 EXPERIMENTAL FRAMEWORK AND RESULTS

The complete experimental work is carried out in two phases. During the first phase, the proposed design is modeled in VHDL. Functional simulation, schematic generation, RTL generation, synthesis for different hardware platforms in FPGA was done and finally implemented in specific target hardware to verify the proper functionality of the design using Xilinx ISE Design Suite 10.1i. In the second phase, VIO core is generated (to be explained in details in the preceeding section) and On-chip verification of the functionality is carried out with the help of Chipscope Pro.

### 3.1 Synthesis and Implementation

The complete design was synthesized and implemented with the use of VHDL using ISE Design Suite 10.1i. Simulation was done by ISE simulator and Modelsim XE simulator. Core generation and on-chip verification was done by Chipscope Pro. Figure 2 represents the different stpes of design during the total course that was followed for the digital implementation. The RTL architecture of DES and TDES is shown in Figure 3 and 4 respectively.

Figure 5 and Figure 6 illustrate the implemented components inside the chip. Additionally, the interconnections of the components are shown.

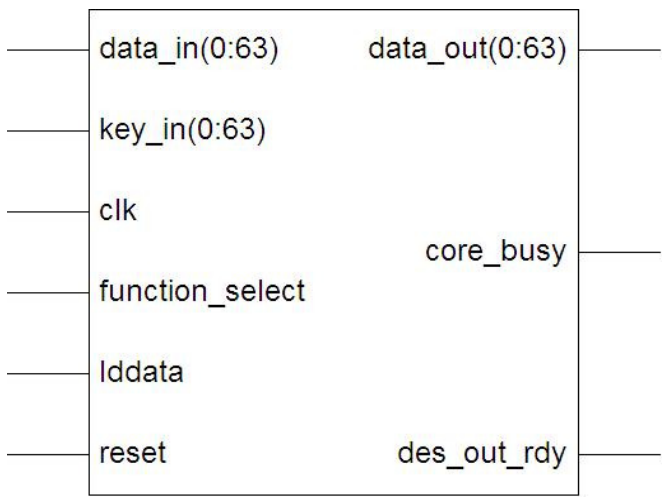

Fig 3: RTL Schematic of DES

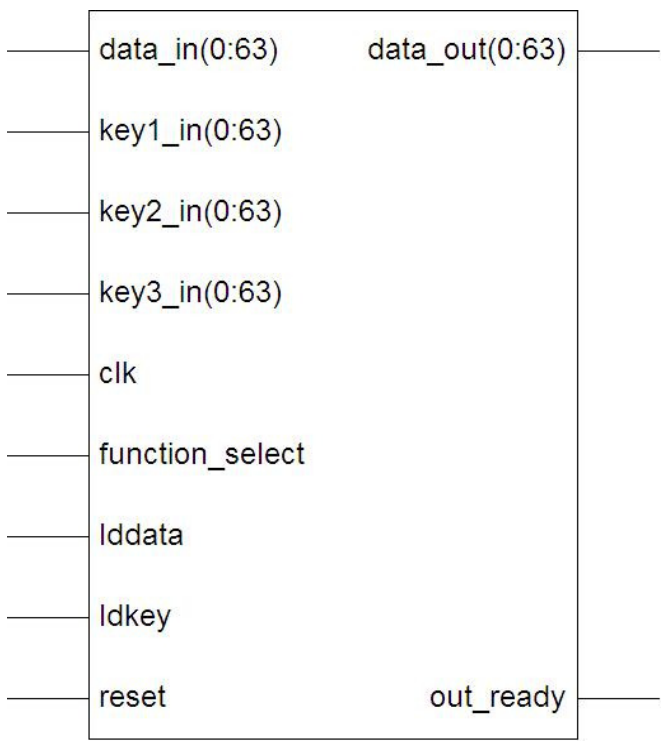

Fig 4: RTL Schematic of TDES

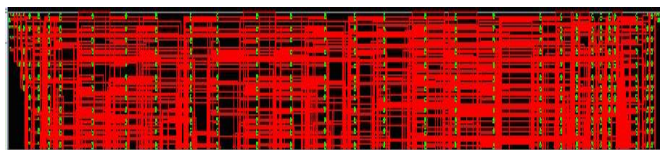

Fig 5: Technology Schematic of DES

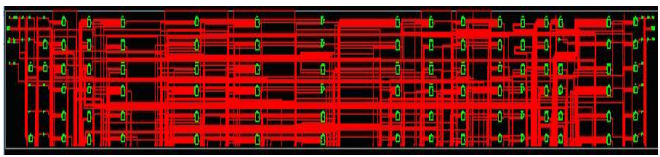

Fig 6: Technology Schematic of TDES

FPGA implementation of DES algorithm and TDES algorithm were accomplished on a Virtex5 device XCVLX5110T using Xilinx ISE Foundation 10.1i. Table 1 and Table 2 show the performance figures for DES hardware implementations. Table 3 and Table 4 show synthesis results of Triple DES implementations.

TABLE 1 (A AND B): DES SYNTHESIS RESULTS FOR SPARTAN SERIES DEVICES

|  | Spartan 3 (Target device xc3s400,Package fg320,Speed -5) | |
|---|---|---|
| Logic Utilization | Used | Utilization |
| Number of Slices | 442 out of 28800 | 5% |
| Number of Slice Flip Flops | 281 out of 28800 | 1% |
| Number of 4 input LUTs | 789 out of 15681 | 5% |
| Number of bonded IOBs | 190 out of 391 | 48% |
| Number of GCLKs | 1 out of 8 | 12% |

|  | Spartan 3AN (Target device xc3s700AN,Package fgg484,Speed -5) | |
|---|---|---|
| Logic Utilization | Used | Utilization |
| Number of Slices | 461 out of 11264 | 4% |
| Number of Slice Flip Flops | 273 out of 22528 | 1% |
| Number of 4 input LUTs | 827 out of 22528 | 3% |
| Number of bonded IOBs | 190 out of 502 | 37% |
| Number of GCLKs | 1 out of 24 | 4% |

TABLE 2 (A AND B): DES SYNTHESIS RESULTS FOR VIRTEX SERIES DEVICES

|  | Vertex 5 (Target device XC5VLX50,Package ff1676,Speed -1) | |
|---|---|---|
| Logic Utilization | Used | Utilization |
| Number of Slice Registers | 266 Out of 28800 | 0% |
| Number of Slice LUTs | 527 Out of 28800 | 1% |
| Number of fully used LUT-FF pairs | 112 Out of 681 | 16% |
| Number of bonded IOBs | 190 Out of 440 | 43% |
| Number of BUFG/BUFGCTRLs | 1 Out of 32 | 3% |

|  | Vertex 5 (Target device XC5VlX110T, Package f1136,Speed -1) | |
|---|---|---|
| Logic Utilization | Used | Utilization |
| Number of Slice Registers | 1,206 Out of 69120 | 1% |
| Number of Slice LUTs | 1,692 OUT OF 69120 | 2% |
| Number of fully used LUT-FF pairs | 632 Out of 2,266 | 27% |
| Number of bonded IOBs | 302 Out of 640 | 47% |
| Number of BUFG/BUFGCTRLs | 1 Out of 32 | 3% |

|  | Vertex 5 (Target device XC5VlX110T, Package f1136,Speed -1) | |
|---|---|---|
| Logic Utilization | Used | Utilization |
| Number of Slice Registers | 266 Out of 69120 | 0% |
| Number of Slice LUTs | 527 Out of 69120 | 0% |
| Number of fully used LUT-FF pairs | 112 Out of 681 | 16% |
| Number of bonded IOBs | 190 Out of 640 | 29% |
| Number of BUFG/BUFGCTRLs | 1 Out of 32 | 3% |

TABLE 3 (A AND B): TRIPLE DES SYNTHESIS RESULTS FOR SPARTAN SERIES DEVICES

|  | Spartan 3 (Target device xc3s1000,Package fg676,Speed -5) | |
|---|---|---|
| Logic Utilization | Used | Utilization |
| Number of Slices | 1585 out of 7680 | 20% |
| Number of Slice Flip Flops | 1254 out of 15360 | 8% |
| Number of 4 input LUTs | 2494 out of 15360 | 16% |
| Number of bonded IOBs | 302 out of 391 | 77% |
| Number of GCLKs | 1 out of 8 | 12% |

|  | Spartan 3AN (Target device xc3s1400AN,Package fgg676,Speed -5) | |
|---|---|---|
| Logic Utilization | Used | Utilization |
| Number of Slices | 1622 out of 11264 | 14% |
| Number of Slice Flip Flops | 1230 out of 22528 | 5% |
| Number of 4 input LUTs | 2593 out of 22528 | 11% |
| Number of bonded IOBs | 302 out of 502 | 37% |
| Number of GCLKs | 1 out of 24 | 4% |

TABLE 4 ( A AND B): TRIPLE DES SYNTHESIS RESULTS FOR VIRTEX SERIES DEVICES

|  | Vertex 5 (Target device XC5VLX50,Package ff1676,Speed -1) | |
|---|---|---|
| Logic Utilization | Used | Utilization |
| Number of Slice Registers | 1206 out of 28800 | 4% |
| Number of Slice LUTs | 1690 out of 28800 | 5% |
| Number of fully used LUT-FF pairs | 447 out of 2449 | 18% |
| Number of bonded IOBs | 302 out of 440 | 68% |
| Number of BUFG/BUFGCTRLs | 1 out of 32 | 3% |

TABLE 5: TRIPLE DES SYNTHESIS RESULT FOR VIRTEX E SERIES DEVICES AS REPORTED IN [7]

|  | Vertex E (Target device XCV1600E,Package bg560,Speed -8) | |
|---|---|---|
| Logic Utilization | Used | Utilization |
| Number of Slices | 12635 out of 14039 | 90% |
| Number of Slice Flip Flops | 20505 out of 31104 | 65% |
| Number of 4 input LUTs | 15518 out of 31104 | 49% |
| Number of bonded IOBs | 243 out of 408 | 59% |
| Number of GCLKs | 1 out of 4 | 25% |

TABLE 6: TRIPLE DES SYNTHESIS RESULT OF OURS

|  | Vertex E (Target device XCV1600E,Package bg560,Speed -8) | |
|---|---|---|
| Logic Utilization | Used | Utilization |
| Number of Slices | 1481 out of 15552 | 9% |
| Number of Slice Flip Flops | 1256 out of 31104 | 4% |
| Number of 4 input LUTs | 2396 out of 31104 | 7% |
| Number of bonded IOBs | 302 out of 404 | 74% |
| Number of GCLKs | 1 out of 4 | 25% |

The comparison between achieved results of DES for vertex E series and the existing implementations for Vertex E series [7] are represented in Table 6 and 5 respectively. It is clear from the comparison result that the proposed implementation is very much compact and efficient in all respects. The synthesis is carried out with the same Virtex E series device resulting in very less consumption of hardware components.

### 3.2 Core Generation, Analysis and On-chip Verification using Chipscope Pro

Xilinx Core Generator Tool provides core generation capability for the Integrated Controller core (ICON), Integrated Logic Analyzer core (ILA), Virtual Input/Output core (VIO), Integrated Bit Error Ratio core (IBERT), Agilent Trace Core 2 (ATC2). ChipScope Pro tools integrate key logic analyzer hardware components with the target design inside the supported devices. The ICON core provides a communication path between the JTAG Boundary Scan port of the target FPGA whereas the VIO core is a customizable core that can both monitor and drive internal FPGA signals in real time.

In this design, after generating programming file three primary steps are:

1. Generate ICON & VIO cores using ChipScope Pro Generator.

2. Analyze & Implement the Design in ISE.
3. Drive & observe design inputs & outputs using ChipScope Pro Analyzer.

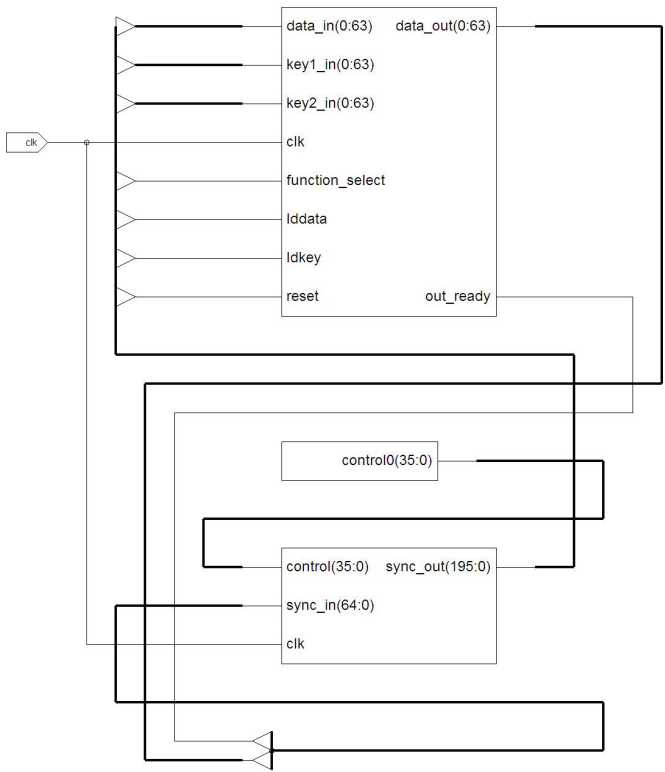

Fig 7: RTL Schematic of the VIO Core of Triple DES using Chipscope Pro

## 4 CONCLUSION

The proposed implementation of DES and TDES provide high-speed performance with very compact hardware implementation. It is a flexible solution for any cryptographic system and security layers of wireless protocol. Measurement results and comparisons between the proposed and previous hardware implementations are presented that shows quite encouraging results.

## ACKNOWLEDGMENT

This research work was supported in part by the grant from All India Council for Technical Education (AICTE) under Research Promotion Scheme (RPS) by grant no. 8023/BOR/RID/RPS-92/07/08 dated 05.03.2008.

**Prasun Ghosal** is curently associated as an Assistant Professor with the Department of Information Technology in Bengal Engineering & Science University, Shibpur. He is presently pursuing for his Doctoral Award and completed his M.Tech. (2005) and B.Tech. (2002) from Institute of Radio Physics & Electronics, University of Calcutta, India. He is also an Honours Graduate (major in Physics) from R. K. Mission Vidyamandira, Belur (1999) under University of Calcutta. His research interests include 3D integration in VLSI Physical Design, Design of Embedded Systems, Network-on-Chip. He is a member of IEEE.

**Malabika Biswas** has completed her ME in Information and Communication Engineering (ICE) from Bengal Engineering and Science University, Shibpur in 2009.

**Manish Biswas** has completed her ME in Information and Communication Engineering (ICE) from Bengal Engineering and Science University, Shibpur in 2009.